\documentclass[sigconf]{acmart}
\renewcommand\footnotetextcopyrightpermission[1]{} 
\renewcommand{\headrulewidth}{0pt} 

\acmSubmissionID{5816}
\AtBeginDocument{%
  }

\setcopyright{rightsretained}
\copyrightyear{2026}
\acmYear{2026}
\acmDOI{XXXXXXX.XXXXXXX}
\acmConference[ACM Multimedia '26]{the 2026 ACM International Conference on Multimedia}{November 10--14, 2026}{Rio de Janeiro, Brazil}
\acmISBN{}

\begin{document}

\title{SandSim: Curve-Guided Gaussian Splatting for Reconstructing Sand Painting Processes}

\author{Yilin Wang}
\affiliation{%
  \institution{School of Computer Science and Technology}
  \state{East China Normal University Shanghai}
  \country{China}
}
\email{10225102434@stu.edu.ecnu.cn}

\author{Haojie Huang}
\affiliation{%
  \institution{School of Computer Science and Technology}
  \state{East China Normal University Shanghai}
  \country{China}
}
\email{51285900013@stu.edu.ecnu.cn}

\author{Chen Li}
\affiliation{%
  \institution{School of Computer Science and Engineering}
  \state{Tianjin University of Technology}
  \country{China}
}
\email{cli@email.tjut.edu.cn}

\author{Yang Li}
\affiliation{%
  \institution{School of Computer Science and Technology}
  \state{East China Normal University Shanghai}
  \country{China}
}
\email{yli@cs.ecnu.edu.cn}

\author{Changbo Wang}
\affiliation{%
  \institution{School of Computer Science and Technology}
  \state{East China Normal University Shanghai}
  \country{China}
}
\email{cbwang@cs.ecnu.edu.cn}

\author{Chenhui Li}
\authornote{Corresponding author}
\affiliation{%
  \institution{School of Computer Science and Technology}
  \state{East China Normal University Shanghai}
  \country{China}
}
\email{chli@cs.ecnu.edu.cn}

\renewcommand{\shortauthors}{Trovato et al.}


\begin{abstract} Sand painting is a process-driven art where visual appearance emerges from granular accumulation. Given a single image, reconstructing a plausible sand painting process requires modeling coherent stroke structures and material-dependent effects. Existing methods, including stroke-based optimization and diffusion-based video synthesis, often lack structural coherence and material consistency, leading to unrealistic drawing sequences. We present SandSim, a framework that reconstructs a sand painting process from a single image. We introduce a curve-guided Gaussian representation that models strokes as sequences of anisotropic primitives along continuous trajectories, whose smooth kernels capture the soft boundaries of sand strokes and enable coherent stroke formation. We further adopt a subtractive compositing scheme to model light attenuation during sand accumulation. We incorporate a semantic-guided planning module for scene decomposition and drawing order inference. Our framework jointly optimizes stroke geometry and appearance and can be integrated with a physics-based simulator for interactive sand dynamics and editing. Experiments show that our method produces temporally coherent and visually realistic results, achieving improved reconstruction quality and perceptual fidelity compared to existing approaches. \end{abstract}


\begin{teaserfigure}
  \includegraphics[width=\textwidth]{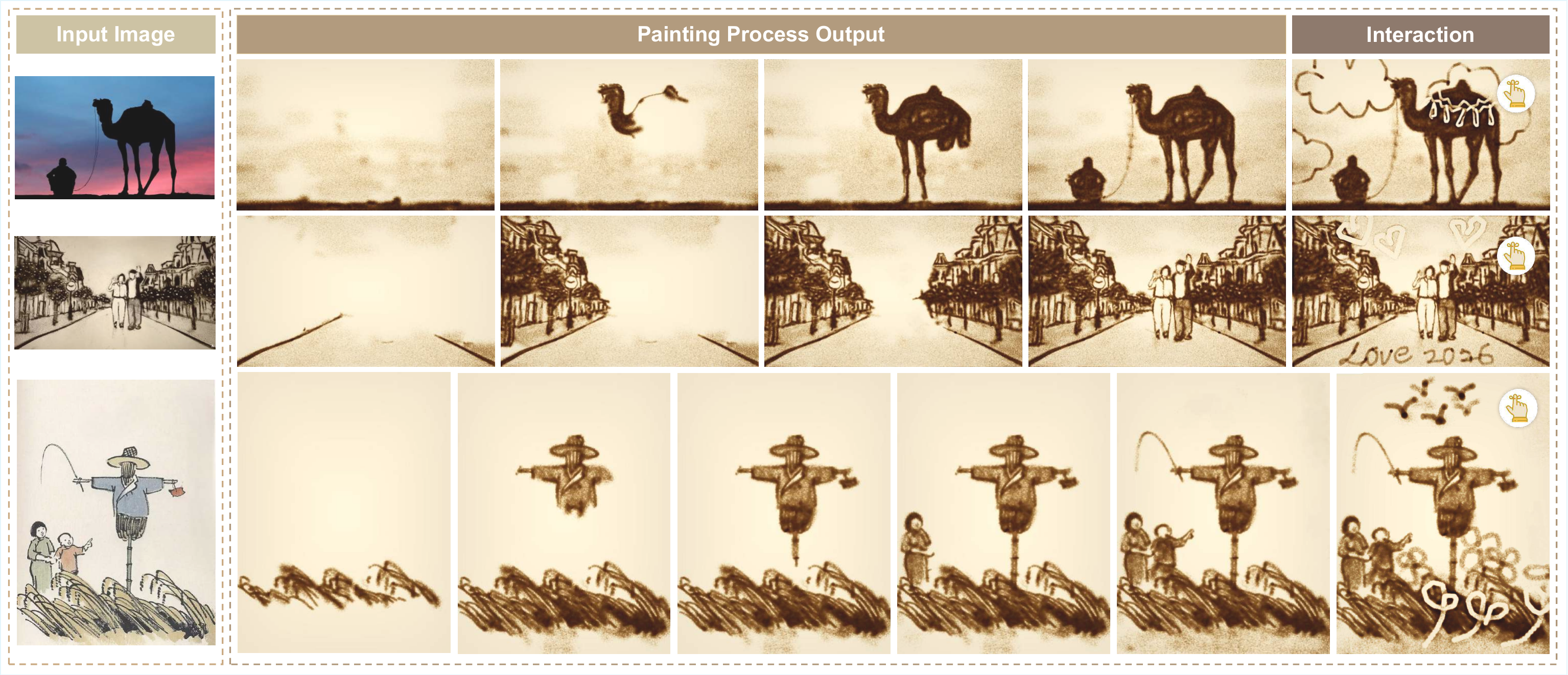}
  \caption{Illustration of our results. Given an input image (left), our method reconstructs a dynamic sand painting process
(middle) and supports interactive editing (right). The first example shows reconstruction from a real photograph, the second
from a real sand painting result, and the third from a sketch input.}
  \label{fig:teaser}
\end{teaserfigure}


\maketitle

\thispagestyle{empty}
\pagestyle{empty}
\fancyhead{}
\fancyfoot{}
\renewcommand{\headrulewidth}{0pt}

\begin{figure*}[t] 
  \centering
  \includegraphics[width=\textwidth]{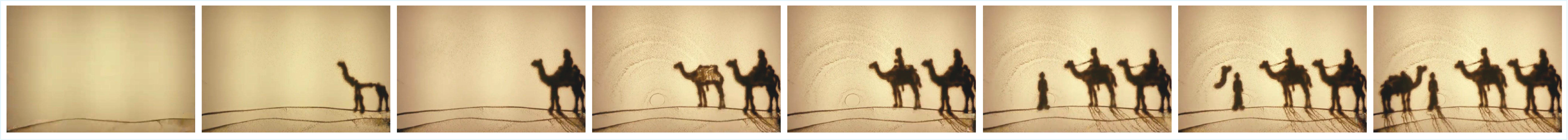} 
  
\caption{A real sand artist's drawing process (captured from YouTube). 
The sequence illustrates a typical back-to-front strategy, where the ground is established first, followed by foreground subjects such as the camels and rider. 
The artist focuses on one semantic region at a time, maintaining structural coherence throughout the process.}
  \label{fig:artist_process}
\end{figure*}

\section{Introduction}
A painting hangs on a wall, while a sand painting unfolds in time. Unlike conventional painting, sand painting is inherently procedural: the final image is the residue of a temporally extended creation process. Before placing any sand, the artist forms a mental plan of the semantic layout and stroke order. The drawing then proceeds stroke by stroke in an iterative manner, typically following a back-to-front strategy, as shown in Fig.~\ref{fig:artist_process}. 
This naturally leads to the problem of recovering the underlying drawing sequence from a single static image. Addressing this problem requires modeling not only visual appearance, but also the geometric, physical, and semantic structure of the painting process.

Existing methods for painting process reconstruction, ranging from stroke-based representations \cite{liu2021paint, huang2020learning, zou2021stylized, hu2024towards, hu2023stroke, nolte2022stroke, tong2022im2oil} to diffusion-based approaches \cite{chen2024inverse, processpainter, zhang2025generating, pobitzer2025loomis}, primarily focus on recovering static frames. In these methods, strokes are often treated as discrete elements, and their temporal evolution is not explicitly modeled. As a result, the continuous nature of stroke formation and the interaction of granular materials are not fully captured.

We formulate the task as a stroke-wise reconstruction problem, where the goal is to generate a sand painting process video from a single image. A natural approach is to use stroke-based representations. However, existing methods \cite{liu2021paint, hu2024towards, hu2023stroke} rely on straight or piecewise linear strokes, which are not well suited for representing curved and continuous structures.
To address this limitation, we introduce a curve-based Gaussian representation, where each stroke is modeled as a sequence of Gaussian primitives distributed along a smooth curve. This representation maintains stroke continuity and provides a flexible way to approximate sand deposition along trajectories.

Based on this representation, we further introduce a semantic planning module that performs semantic region decomposition, drawing order inference, and stroke-level classification. This allows the generation of structured drawing sequences that align with both high-level semantics and low-level stroke formation, while preserving temporal consistency across the reconstructed process.
We also develop an interactive sand painting system built on the same Gaussian representation, enabling real-time sand deposition and manipulation during both the drawing process and the resulting image, as shown in Fig.~\ref{fig:teaser}.

In summary, our main contributions are as follows:
\begin{itemize}
    \item \textbf{Curve-guided Gaussian Representation.}  
    We design a parametric curve-guided Gaussian representation that models strokes as continuous sequences of anisotropic primitives, enabling coherent stroke formation and high-fidelity reconstruction via Gaussian splatting.

    \item \textbf{Physics-aware Sand Painting Modeling.}  
We model sand painting as a continuous physics-based process using Gaussian strokes to simulate sand deposition and accumulation, and introduce a subtractive compositing scheme to better capture light attenuation in granular materials, improving realism over additive rendering.

    \item \textbf{Semantic Planning for Drawing Sequences.}  
    We incorporate a semantic planning module for scene decomposition, drawing order inference, and stroke-level structuring, enabling more coherent and structured drawing processes.
\end{itemize}

\begin{figure*}[t] 
  \centering
  \includegraphics[width=\textwidth]{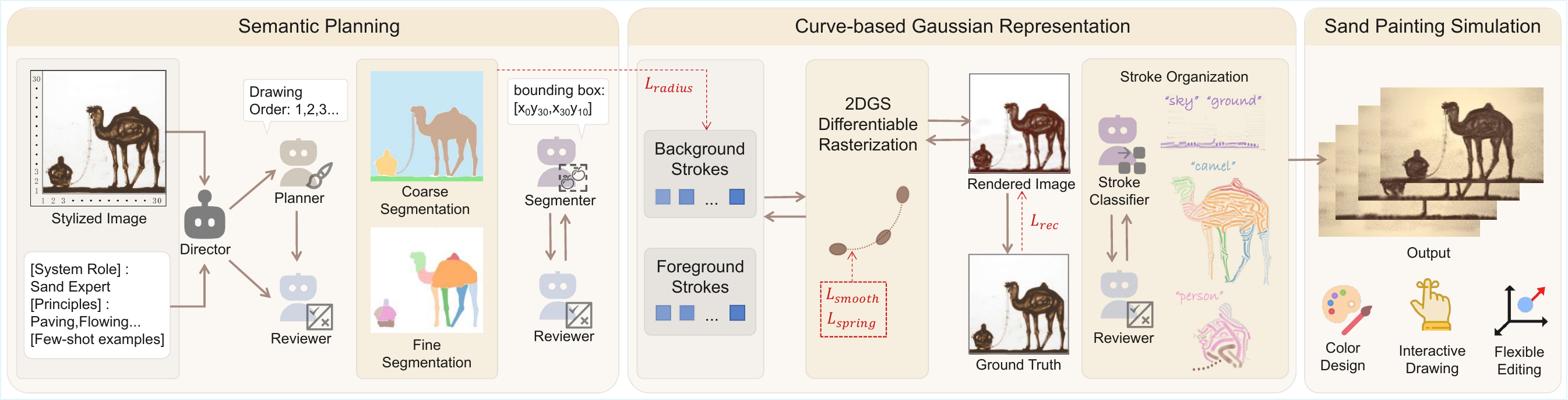} 
  
\caption{\textbf{Overview of our framework.} 
The pipeline begins with a semantic planning module, which decomposes the image and infers a logical drawing order. Next, the curve-based Gaussian representation parameterizes strokes as smooth curves. Finally, the sand painting simulation physically instantiates the strokes to generate a dynamic, step-by-step painting process, enabling interactive editing.}

  \label{fig:pipeline}
\end{figure*}

\section{Related Work}

\subsection{Painting Process Generation}

Painting process generation has been predominantly studied under the framework of Stroke-Based Rendering (SBR), which formulates painting as a sequential stroke optimization problem. Prior works mainly focus on generating stylized static images \cite{kotovenko2021rethinking, liu2023painterly, zou2021stylized}, often overlooking the temporal order and dynamics of stroke placement.

Subsequent approaches aim to produce more human-like processes by modeling drawing trajectories with parameterized strokes. Reinforcement learning is widely used for sequential stroke optimization \cite{de2024segmentation, hu2023stroke, huang2020learning, singh2021combining, zou2021stylized}, while Transformer-based methods \cite{vaswani2017attention} enable coarse-to-fine stroke prediction \cite{liu2021paint}. Other works incorporate progressive stroke layering \cite{singh2022intellipaint}, semantic guidance and agent-based frameworks \cite{hu2024towards, vinker2025sketchagent, wang2024genartist}, or novel stroke representations \cite{qin2024hyperstroke}. Despite these advances, strokes are still often modeled as discrete primitives, limiting the modeling of continuous stroke evolution.

Recent works explore diffusion-based approaches to model painting processes via generative video modeling. These methods learn drawing dynamics from data and generate intermediate frames sequentially. For example, Inverse Painting \cite{chen2024inverse} reconstructs drawing sequences autoregressively, while ProcessPainter \cite{processpainter}, Generating Past \cite{zhang2025generating}, and Loomis \cite{pobitzer2025loomis} leverage video diffusion priors to produce plausible painting processes.

However, existing methods have limitations. Diffusion-based approaches model the painting process as a sequence of images, often suffering from limited temporal consistency. In addition, they lack an explicit representation of the physical space. SBR-based methods rely on parameterized strokes to generate smooth trajectories, which limits their ability to capture complex variations.
In contrast, we represent the painting process as a continuous sequence of stroke-level operations and render it using a physics-based sand renderer, enabling temporally consistent generation while maintaining realistic appearance.

\subsection{Gaussian Splatting}
3D Gaussian Splatting (3DGS) \cite{kerbl2023gaussiansplatting} revolutionizes novel view synthesis via explicit point-based primitives and tile-based rasterization, bypassing costly volumetric ray-marching for real-time photorealistic rendering, and extends to spatiotemporal dynamics \cite{wu2024gaussiansplatting4d}. Beyond 3D scenes, 2D Gaussians enable efficient image representation and compression \cite{zeng2025instantgaussianimage, zhang2024gaussianimage, zhu2025largeimagesgaussians}. To connect continuous geometry, Bézier Splatting \cite{liu2025beziersplatting, gao2025curve} and BG-Triangle \cite{wu2025bgtriangle} integrate Gaussians with parametric vector curves for differentiable SVG rendering. Explicit Gaussian parameterization also benefits artistic editing and physical actuation. While some methods target global style transfer or static deformations \cite{liu2024stylegaussian, zhang2025stylizedgs, fang2025a3gs, wynn2025morpheus}, interactive systems employ Gaussians as 3D brushstrokes \cite{pandey2025gaussianbrush}. For material dynamics, PhysGaussian \cite{xie2024physgaussian} and DreamPhysics \cite{huang2025dreamphysics} embed continuum mechanics into Gaussian kernels. Nevertheless, most frameworks still treat Gaussians either as static visual proxies or low-level physical surrogates, lacking a unified approach for high-fidelity visual reconstruction guided by top-down semantic drawing.

\subsection{Sand Simulation}
Simulating granular materials like sand is challenging due to their solid-fluid duality. Early approaches, including pure fluid approximations, discrete element methods (DEM), and 2.5D height-fields \cite{zhu2005animating, bell2005particle, zhu2019shallow}, are efficient but struggle to capture complex 3D topological changes. The Material Point Method (MPM) overcomes this by tracking Lagrangian particles on an Eulerian grid, becoming the standard for various complex materials \cite{stomakhin2013material, jiang2015affine, jiang2017anisotropic}. Its performance has been massively accelerated by GPU architectures, Moving Least Squares (MLS), and Taichi optimizations \cite{gao2018gpu, hu2018moving, hu2019taichi}. For sand specifically, integrating Drucker-Prager elastoplasticity \cite{klar2016drucker} accurately models frictional yield. This foundation, combined with recent advances in rigid-body coupling, compact-kernel designs, and complex sand-water mixtures \cite{daviet2021frictional, liu2025ck, tang2025granule}, makes MPM the gold standard for granular physics. However, traditional rendering of millions of particles remains memory-intensive and non-differentiable \cite{meng2015multiscale, liu2025sandtouch}. Recent methods like PhysGaussian \cite{xie2024physgaussian} address this by coupling MPM with 3DGS, seamlessly overcoming the rendering bottleneck while preserving high-fidelity physical realism.

\section{Method}
Given an input image, our goal is to reconstruct a sand painting process that progressively adds sand strokes to reproduce the target image. Inspired by real sand painting creation, we apply LoRA-based style transfer to obtain a stylized input, then propose a synthesis pipeline integrating semantic planning, parametric stroke representation, and physical–based sand rendering. Fig.~\ref{fig:pipeline} shows an overview of the method.

\subsection{Semantic Planning}

This module aims to perform structured analysis of the input image and decompose it into semantic regions aligned with the logic of sand painting. Unlike conventional image segmentation methods that focus on boundary detection, it jointly reasons about artistic composition and drawing order. As illustrated in the Semantic Planning module of Fig.~\ref{fig:pipeline}, we decouple the cognitive process of sand painting into five collaborative agents. Built upon a unified conversational interface, these agents interact through structured message passing and shared context, forming a dialogue-driven control flow to complete the planning process.

\textbf{Director.} This agent initializes the system’s artistic understanding via a comprehensive few-shot prompting strategy. By configuring the system message, it establishes an expert role in sand painting, introduces the fundamental principles of sand painting, and provides few-shot examples of drawing sequences to guide the model in aligning visual elements with domain knowledge.

\textbf{Planner.} Guided by the artistic principles initialized by the Director, this agent determines the temporal drawing sequence of objects. It analyzes semantic regions and assigns core attributes such as label, layer, and painting method, and evaluates whether fine-grained segmentation is required. For structurally complex objects that require further refinement, it further lists their semantic sub-components, for example, head, torso, and arms, for stage-two processing.

\textbf{Segmenter.} This agent performs semantic segmentation based on the directives from the Planner. It acts as a tool-execution proxy, bridging language reasoning and vision models. To overcome the inherent spatial perception limitations, it overlays a $30 \times 30$ grid on the input image. Using grid coordinates, it predicts precise bounding boxes for the coarse regions and fine-grained parts specified by the Planner, for example, represented by coordinate pairs $(x_i, y_j)$. These predicted bounding boxes, together with their corresponding labels, serve as visual prompts to trigger the Segment Anything Model (SAM)~\cite{kirillov2023segment} to generate segmentation masks.

\textbf{Stroke Classifier.} As a post-processing module following the stroke fitting stage (Sec.~\ref{sec:training}), it performs micro-level optimization. Its primary role is to analyze and re-classify all drawing strokes that are not explicitly covered by the SAM-generated masks, assigning them to the correct semantic regions based on their morphological features and the semantic context they belong to.

\textbf{Reviewer.} This agent implements a closed-loop feedback mechanism for the planning process. It monitors each stage of the pipeline and evaluates the outputs from preceding agents with respect to the artistic logic of sand painting. When spatial, stylistic, or logical inconsistencies are identified, it generates targeted feedback and feeds it back into the conversational process. This feedback prompts the corresponding agent to update its output, enabling iterative refinement of the overall plan.

\subsection{Primitives of SandSim}
\label{sec:primitives}

\begin{figure}[t]
  \centering
  \includegraphics[width=\linewidth]{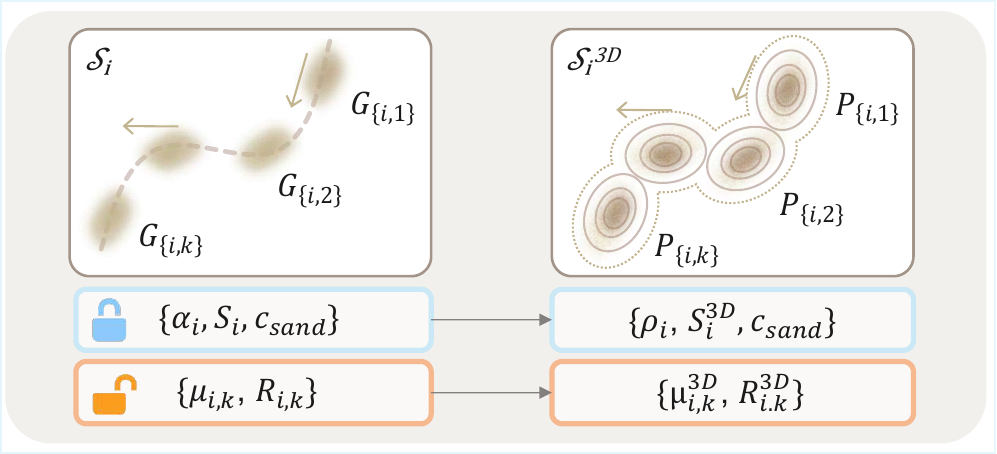}
  \caption{Illustration of the 2D-to-3D stroke lifting process. 
The hierarchical attributes of 2D Gaussian strokes are mapped to 3D Gaussian particle clusters.
}
  \label{fig:stroke_lifting} 
\end{figure}

We represent a sand painting as a collection of curve-guided Gaussian sand strokes. As illustrated in the left panel of Fig. \ref{fig:stroke_lifting}, Our representation utilizes a sequence of 2D Gaussian kernels to inherently mimic the physical process of sand particles falling along a drawing trajectory.

\textbf{Gaussian sand kernels. }The fundamental primitive in our framework is the 2D Gaussian kernel. A sand stroke $S_i$ is represented as a sequence of Gaussian kernels $\{G_{i,k}\}$, where each kernel $G_{i,k}$ is characterized by its 2D center position $\mu_{i,k} \in \mathbb{R}^2$, an anisotropic covariance matrix $\Sigma_{i,k}$, a color vector $c_{\text{sand}} \in \mathbb{R}^3$, and a shared opacity $\alpha_{i} \in [0,1]$. The kernel function is defined as:\begin{equation}G_{i,k}(x) = \alpha_{i}, \exp\Big(-\tfrac{1}{2} (x - \mu_{i,k})^{\top} \Sigma_{i,k}^{-1} (x - \mu_{i,k}) \Big).\end{equation}The covariance matrix $\Sigma_{i,k}$ encodes the anisotropic shape and orientation of the kernel. To facilitate stroke-level parameter sharing while maintaining optimization flexibility, we decompose the covariance matrix into a scaling matrix $S_i$ and a rotation matrix $R_{i,k}$:\begin{equation}\Sigma_{i,k} = R_{i,k} S_i S_i^{\top} R_{i,k}^{\top},\end{equation}where $S_i = \text{diag}(s_{i,x}, s_{i,y})$ is a diagonal matrix representing the scaling factors along the principal axes.In our curve-guided representation, to maintain morphological consistency, the scaling matrix $S_i$ and opacity $\alpha_i$ are shared across all kernels $\{G_{i,k}\}$ within the same stroke $S_i$, defining its uniform thickness and sand density. Conversely, the rotation matrix $R_{i,k}$ is optimized as a per-point parameter, enabling each kernel to dynamically align with the local tangent of the drawing trajectory. The color $c_{\text{sand}}$ serves as a fixed subtractive absorption coefficient, modeling the physical light attenuation of sand particles.

\textbf{Curve-Guided Gaussian Strokes. }A complete sand painting is formulated as a set of stroke primitives $\{ \mathcal{S}_i \}_{i=1}^{N}
$, where $N$ is the total number of strokes. Each stroke $\mathcal{S}_i$ aggregates a sequence of $K_i$ learnable Gaussian primitives:
\begin{equation}
\mathcal{S}_i = \Big\{ G_{i,k} \mid \mu_{i,k} \in \mathbb{R}^2,\; k = 0, \dots, K_i-1 \Big\}.
\end{equation}
While parametric representations such as Bézier curves \cite{gao2025curve, liu2025beziersplatting} produce smooth results 
when modeling regular geometric edges, they are less suitable for representing 
the irregular filling structures commonly observed in sand painting strokes.
To maintain smooth and continuous drawing trajectories while producing more natural stroke patterns, we introduce a geometric regularization loss $\mathcal{L}_{\text{geom}}$. The first component is the spring loss $\mathcal{L}_{\text{spring}}$, which models each stroke as a discrete mass-spring system. In this analogy, each Gaussian center $\mu_{i,k}$ acts as a physical particle connected to its neighbors by virtual springs:
\begin{equation}
\mathcal{L}_{\text{spring}} = \sum_{k=0}^{K_i-2} \|\mu_{i,k+1} - \mu_{i,k}\|^2,
\end{equation}
This term exerts a cohesive restoring force that encourages neighboring centers to maintain a uniform spacing, preventing the stroke from scattering into isolated dots. The second component is the smoothness loss $\mathcal{L}_{\text{geom}}$, which penalizes the bending energy of the sequence to enforce local linearity:
\begin{equation}
\mathcal{L}_{\text{smooth}} = \sum_{k=1}^{K_i-2} \|\mu_{i,k-1} - 2 \mu_{i,k} + \mu_{i,k+1}\|^2.
\end{equation}
This discrete Laplacian operator suppresses high-frequency geometric noise, resulting in visually fluid and graceful curves. The complete geometric regularization is defined as:
\begin{equation}
\mathcal{L}_{\text{geom}} = \lambda_{\text{spring}} \mathcal{L}_{\text{spring}} + \lambda_{\text{smooth}} \mathcal{L}_{\text{smooth}}.
\end{equation}
Here, $\lambda_{\text{spring}}$ and $\lambda_{\text{smooth}}$ are hyper-parameters assigned based on the stroke length to balance structural integrity.

\subsection{Differentiable Rasterization}
\label{sec:rasterization}

To simulate the visual appearance of sand painting during optimization, we adopt a differentiable rasterization pipeline based on 2D Gaussian Splatting, with modifications tailored to the optical characteristics of sand.

\textbf{Subtractive Color Model.}
In sand painting, visual appearance is governed by progressive light absorption: as sand accumulates, the background is increasingly occluded, resulting in darker intensities.
To model this behavior, we represent the canvas as a white background $\mathbf{b}$ and use a fixed subtractive absorption coefficient $c_{\text{sand}}$, as defined in Sec.~\ref{sec:primitives}. This formulation avoids the unrealistic brightness accumulation produced by additive blending in overlapping regions.
Furthermore, since single-image sand painting lacks explicit depth information, we adopt an order-independent accumulation scheme that directly aggregates the contributions of all Gaussian primitives at each pixel, without depth sorting.

The rendered color $C_{\text{train}}(x)$ at pixel $x$ is defined as:
\begin{equation}
    C_{\text{train}}(x) = \max\!\left(
    \mathbf{b} - c_{\text{sand}}
    \sum_{i=1}^{N} \sum_{k=0}^{K_i - 1} G_{i,k}(x),
    \; \mathbf{0}
    \right),
    \label{eq:train_render}
\end{equation}
This formulation is a first-order approximation of the Beer-Lambert(GLCM 
absorption law, which is accurate for typical sand painting scenarios 
where layer thickness remains moderate.

\textbf{Differentiable pipeline.}
This formulation eliminates the need for depth sorting while remaining fully differentiable, allowing gradients to effectively optimize the Gaussian parameters, including position $\mu_{i,k}$, covariance $\Sigma_{i,k}$, and opacity $\alpha_i$.

\subsection{Training and Optimization}
\label{sec:training}

\textbf{Loss design.}
Given an input image $\mathbf{I} \in \mathbb{R}^{H \times W \times 3}$, SandSim aims to vectorize it into a collection of curve-guided Gaussian sand strokes while ensuring high-fidelity reconstruction. Stroke primitives are randomly initialized on the canvas and rendered via the differentiable 2DGS pipeline, producing $\hat{\mathbf{I}} \in \mathbb{R}^{H \times W \times 3}$. All stroke parameters are optimized end-to-end using gradient-based loss functions.

\paragraph{Reconstruction loss.}
Following GaussianImage~\cite{zhang2024gaussianimage}, we optimize the Gaussian parameters
by minimizing the reconstruction error between the rendered image
$\hat{I}$ and the target image $I$ using an $\ell_2$ loss:
\begin{equation}
    \mathcal{L}_{\text{rec}} = \|\hat{I} - I\|_2^2.
\end{equation}

\paragraph{Geometric regularization.}
As defined in Sec.~\ref{sec:primitives}, the geometric regularization $\mathcal{L}_{\text{geom}}$ enforces stroke smoothness and cohesion by combining spring and smoothness losses. This ensures that freely optimizable Gaussian centers form visually coherent and physically plausible drawing trajectories.

\paragraph{Region-aware scale regularization.}
To simulate the inherent layered structure of sand painting, we apply different constraints to foreground and background regions. Based on the region partition obtained from Sec.~3.1, we classify strokes according to their spatial overlap with these regions. For background regions, we encourage strokes to adopt larger scales to mimic the physical behavior of coarse sand accumulation:

\begin{equation}
\mathcal{L}_{\text{scale}} = \sum_{i \in \mathcal{S}_{\text{bg}}} \|\mathbf{s}_i - \mathbf{s}_{\text{target}}\|_2^2,
\label{eq:scale_loss}
\end{equation}
where $\mathcal{S}_{\text{bg}}$ denotes the set of background strokes, $\mathbf{s}_i \in \mathbb{R}^2$ is the learnable scale vector of the $i$-th Gaussian stroke, and $\mathbf{s}_{\text{target}}$ is the predefined target scale vector. The total training loss combines all three components:
\begin{equation}
    \mathcal{L} = \mathcal{L}_{\text{rec}} + \lambda_g \mathcal{L}_{\text{geom}} + \lambda_s \mathcal{L}_{\text{scale}},
    \label{eq:total_loss_training}
\end{equation}
where $\lambda_g$ and $\lambda_s$ serve as hyper-parameters, balancing geometric smoothness and region-aware constraints. 

\textbf{Dynamic topology.}
During optimization, the stroke topology is dynamically adjusted to balance reconstruction quality and representation efficiency. Specifically, \textbf{point merging} removes redundant Gaussian centers along a stroke to maintain uniform spacing; \textbf{stroke merging} combines strokes with aligned endpoints and similar attributes into longer or thicker strokes; \textbf{stroke splitting} divides strokes when adjacent Gaussian centers become too sparse, preventing cross-region fitting; and \textbf{stroke pruning} removes weak or insignificant strokes, with additional pruning applied in later stages.These topology operations are performed periodically throughout training, ensuring that the primitive set remains compact while sufficiently capturing complex image details.

\subsection{Physical Sand Painting Simulation}
\label{sec:simulation}

While the 2D vectorization (Sec.~\ref{sec:training}) produces optimized 
stroke trajectories, real sand paintings exhibit three-dimensional granular 
structures formed by physical particle accumulation. To bridge this gap, we 
introduce a physics-based simulation stage that lifts the optimized 2D strokes 
into physically plausible 3D sand structures.

\textbf{2D-to-3D stroke lifting.}
To physically instantiate the optimized 2D trajectories, we formulate a 
parameter-consistent 2D-to-3D lifting scheme, as illustrated in Fig.~\ref{fig:stroke_lifting}.

\textit{Stroke-level mapping.}
Given a 2D Gaussian stroke represented by $\{\alpha_i, s_i, c_{\text{sand}}\}$, 
we construct its 3D counterpart $\{\rho_i, s_i^{3D}, c_{\text{sand}}\}$, where
\begin{equation}
    \rho_i = f(\alpha_i) 
    = \frac{1 - \exp(-\alpha_i)}{1 - \exp(-1)}.
\end{equation}
Here, the geometric scale and material color are directly inherited, 
while the opacity $\alpha_i$ is converted into a sampling density $\rho_i$.

\textit{Primitive-level mapping.}
Each 2D Gaussian primitive $G_{i,k}$ with parameters 
$\{\mu_{i,k}, R_{i,k}\}$ is lifted to a 3D Gaussian cluster 
$P_{i,k}$ with $\{\mu_{i,k}^{3D}, R_{i,k}^{3D}\}$. 
A set of 3D Gaussians is sampled from a localized distribution centered at $\mu_{i,k}^{3D}$, whose union defines the volumetric sand structure.

\textbf{Granular dynamics and interaction.}  
We adopt the MPM framework from PhysGS~\cite{xie2024physgaussian} with Drucker--Prager plasticity 
to model the granular behavior of sand. Under gravity, particles accumulate 
to form sand piles with varying heights and spatial distributions.
To enable interactive manipulation, we design specialized MPM kernels. 
In particular, to approximate hand contact, the \emph{smear kernel} applies forces within 
a localized interaction region using a paraboloid pressure profile:

\begin{equation}
    p(r) = \max\left(1 - \left(\frac{r}{R}\right)^2, 0\right)
\end{equation}
This formulation produces stronger displacement near the center and a smooth falloff toward the boundary.

A key issue in interactive MPM is unintended momentum propagation through the grid. 
To address this, we introduce a freeze mechanism that selectively constrains 
particle velocities outside the interaction region. Specifically, particles beyond 
a safe radius $R_{\text{safe}}$ are clamped when their velocity falls below 
a threshold, while fast-moving particles remain unaffected. The final 3D rendering remains consistent with the 2D formulation (Sec.~\ref{sec:rasterization}), 
adopting the same Beer--Lambert \cite{beer1852bestimmung} absorption and subtractive color model.

\section{Experiments}

\subsection{Experimental Setups}
\textbf{Implementation details.}
We implement SandSim in PyTorch and optimize it for 10,000 iterations using the Adan optimizer with a StepLR scheduler where the base learning rate is initialized to 0.005. The model is initialized with 700 curves, each consisting of 20 Gaussian points uniformly distributed across the image. During optimization, we periodically apply dynamic topology operations: adjacent points closer than 0.1 pixels are merged; curves with aligned endpoints within 5 pixels are merged; curves with opacity below 0.01 or radius below 3.0 are pruned; and splitting is triggered when point spacing exceeds the long axis. These operations are disabled during the final 1,000 iterations to stabilize the representation. To enforce visually coherent strokes, we apply spring and smoothness constraints with weights of 0.01 and 0.3, respectively. Region-aware regularizations are applied: background regions receive large-radius regularization with target radius 15.0 pixels and weight 0.1, along with horizontal orientation regularization with weight 1.0, while foreground regions allow elongated ellipses with no radius constraint. Experiments are conducted on a single NVIDIA RTX 3090 GPU.

\textbf{Datasets.}
Our dataset consists of two components: an image dataset for style modeling and a video dataset for capturing the dynamic sand painting process. 
For style learning, we collect 70 high-quality sand painting images covering diverse categories such as portraits, animals, and landscapes. These images are used to train a style-specific LoRA model.
For evaluation, we conduct experiments on our collected sand painting video dataset, which consists of 31 full-length videos from YouTube capturing complete drawing procedures. To alleviate data scarcity and improve diversity, we further construct a synthetic dataset by generating 15 sand painting process videos using the trained LoRA model. The generation is conditioned on images from multiple drawing domains, including line drawings, ink paintings, and sketches, providing diverse structural priors. These real and synthetic data sources enhance the model's ability to generalize across different contents and painting strategies.

\begin{figure*}[t] 
  \centering
  \includegraphics[width=\textwidth]{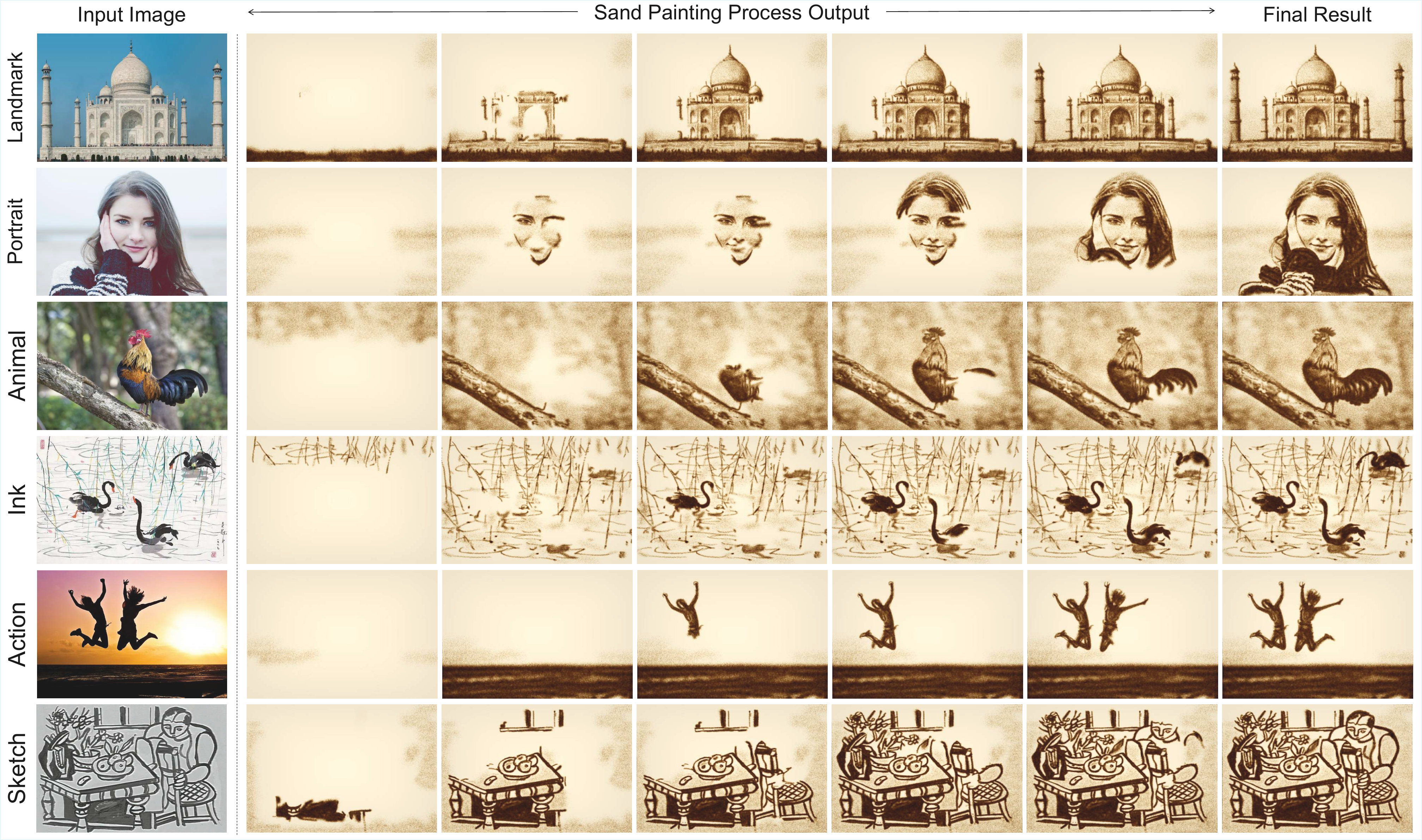} 
  
\caption{Qualitative results.
We present results on diverse image categories (left), while the middle columns show sampled keyframes from the generated sand painting process and the rightmost column shows the final result. 
Our method effectively handles a wide range of visual domains, including landmarks, portraits, and animals, among others.
The generated sequences demonstrate human-like drawing behaviors, including coherent semantic planning, reasonable drawing order, and progressive coarse-to-fine refinement.}

  \label{fig:quali}
\end{figure*}

\subsection{Qualitative Results}
Fig.~\ref{fig:quali} shows the results of our method on various images. 
First, our method generalizes well across diverse visual domains, including images containing animals (rows 3 and 4), images containing humans (rows 2, 5, and 6), and scene images (rows 1, 3, and 4). Second, the generated sand painting process follows human-like drawing principles, progressing from background to foreground, from coarse structures to fine details, and from top to bottom. For example, in row 1, the structure of the building is first roughly established before architectural details are gradually refined. Similarly, in row 5, the global layout of the scene is formed early, followed by the progressive addition of foreground objects. Finally, the generated strokes are coherent and continuous, while also reflecting realistic sand-specific physical properties. For instance, in row 3, the rooster’s tail is gradually extended without abrupt changes, showing strong temporal consistency. In row 2, the background is first rendered with light sand, followed by denser sand in the foreground, illustrating opacity buildup and layering. These results indicate that our method produces smooth, physically plausible sand painting with realistic texture and depth variations.


\subsection{Metrics}
Inspired by \cite{jiang2025ppjudge, chen2024inverse}, 
we evaluate our method from four aspects:
(1) Style and Texture Fidelity. Evaluates the consistency of overall style and the preservation of sand granularity.
(2) Human-like Drawing Order. Assesses how closely the generated sequence follows a plausible human drawing order.
(3) Convergence Speed to Target Artwork. Quantifies how efficiently the process converges toward the target image.
(4) Video Quality. Reflects the overall visual realism and naturalness of the generated video.
Previous work \cite{chen2024inverse, de2024segmentation, huang2020learning, liu2021paint} has proposed some commonly used metrics in the field of image generation. We also adopted the following metrics based on the evaluation method of InversePainting \cite{chen2024inverse}. To further evaluate the granular texture of sand painting, we additionally design a texture quality metric based on GLCM \cite{haralick2007textural}.

LPIPS~\cite{zhang2018unreasonable} and SSIM~\cite{wang2004image} evaluate perceptual similarity and structural consistency between the generated results and the target reference. 
FID~\cite{heusel2017gans} calculates the distance between feature distributions of generated and ground-truth textures to assess overall visual fidelity and realism. 
DDC~\cite{chen2024inverse} measures the discrepancy between LPIPS curves of the generated sequence and the reference process using Dynamic Time Warping (DTW)~\cite{muller2007dynamic}, reflecting convergence dynamics and painting rhythm. 
GTC (Granular Texture Quality) measures sand granularity based on contrast and entropy from the Gray-Level Co-occurrence Matrix (GLCM)~\cite{haralick2007textural}, capturing structural texture fidelity independently of color variations.

Among these metrics, LPIPS and SSIM are used to evaluate aspects (1) and (2). FID addresses aspect (4). DDC serves to assess aspect (3). GTC addresses aspects (1) and (4). Metrics are computed per sample and averaged across the evaluation set. For video data, metrics are computed and averaged over all frames.

\begin{figure}[t]
  \centering
  \includegraphics[width=\linewidth]{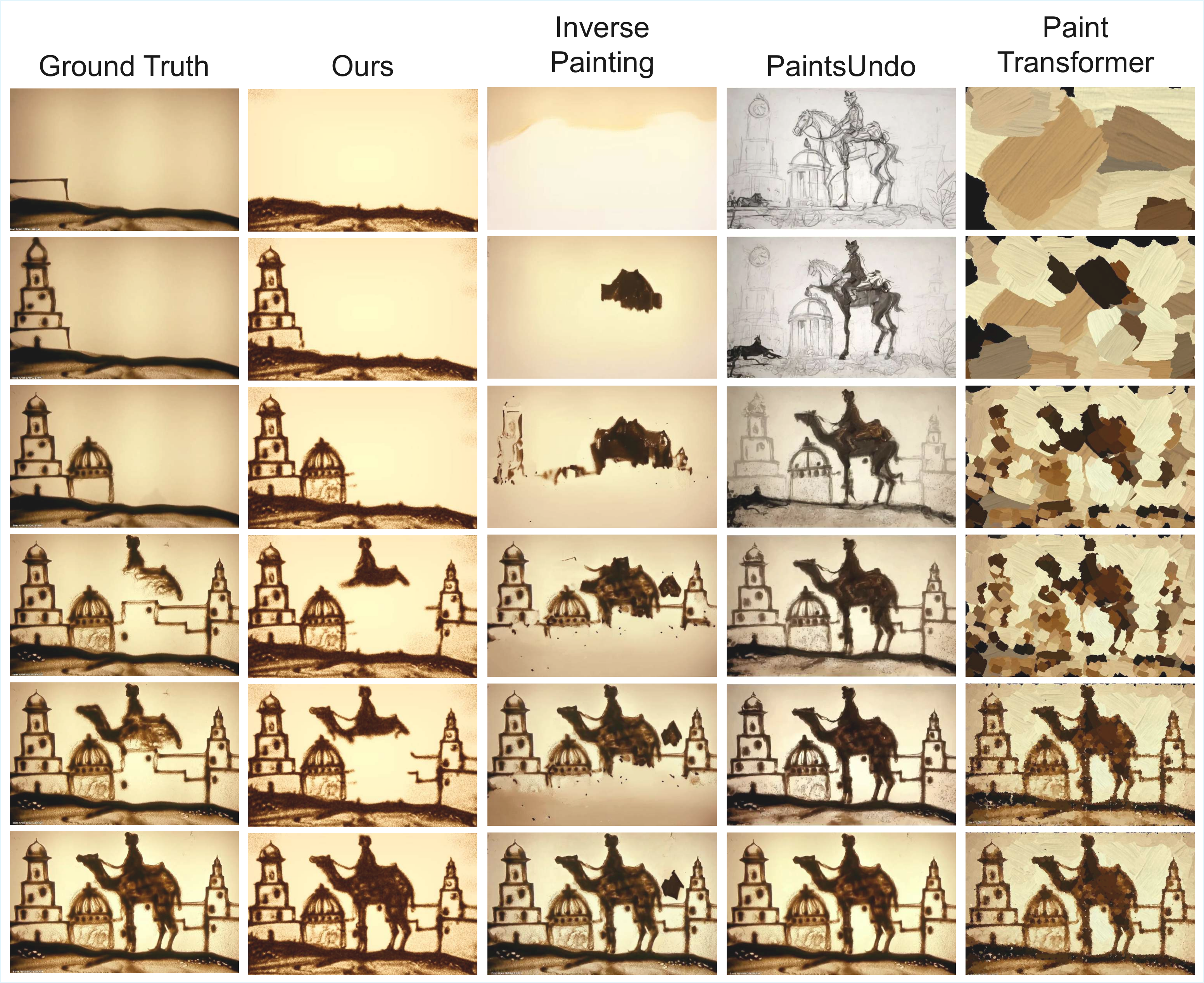}
  \caption{Comparison with baselines. Each column shows sampled keyframes from generated sand painting processes, with the leftmost row as ground truth. Our method produces more human-like and realistic drawing sequences compared to prior approaches.}
  \label{fig:comp}
\end{figure}

\subsection{Comparison with Baselines}
Since our method is designed for sand painting process modeling and simulation, and existing approaches are not tailored to this setting, direct counterparts are not available. In this section, we compare our method with representative state-of-the-art approaches for image-to-drawing-process generation. We select three baselines, including Inverse Painting~\cite{chen2024inverse}, PaintsUndo~\cite{paintsundo}, and Paint Transformer~\cite{liu2021paint}.

\textbf{Qualitative Evaluation.} As shown in Fig.~\ref{fig:comp}, Inverse Painting produces incomplete structures in the background architecture and introduces noticeable artifacts in the final results. PaintsUndo generates drawing sequences that resemble digital painting workflows rather than sand painting, lacking the characteristic accumulation and redistribution patterns of sand. Paint Transformer exhibits a non-human-like drawing order, where strokes are added in parallel across different grid regions of the canvas, resulting in unrealistic process dynamics.
In contrast, our method produces more coherent and physically plausible drawing sequences, better reflecting human-like drawing behaviors and the unique properties of sand manipulation.

\textbf{Quantitative Evaluation.} Table \ref{tab:quantitative_evaluation} reports our quantitative results. Our approach achieves competitive performance across most metrics. Specifically, our method obtains strong LPIPS and SSIM scores, indicating good macroscopic visual quality and reasonable structural consistency with the target image. Furthermore, our DDC score is consistently higher than those of baselines such as Paints-Undo and Paint Transformer, suggesting that our model better captures natural, human-like drawing dynamics.
For sand painting texture fidelity, our method achieves a relatively low GTC. While our FID score is higher than that of Inverse Painting, this may be influenced by a domain gap between sand painting and natural image distributions, as standard FID may not fully align with the characteristics of scattered sand patterns. This observation suggests that traditional metrics alone may not fully reflect the physical granularity of generated sand, motivating the use of our proposed GTC metric as a complementary evaluation.

\begin{figure}[t]
  \centering
  \includegraphics[width=\linewidth]{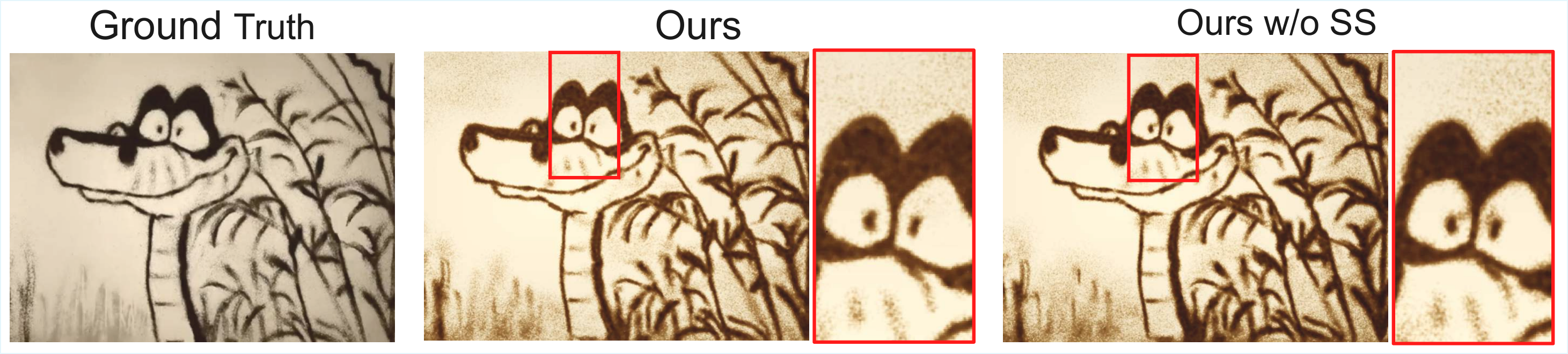}
  \caption{\textbf{Ablation of Stroke Splitting.} 
Without stroke splitting, primitives cross semantic regions and merge fine details such as the snake's eyes. 
Our full method preserves clear boundaries.}

  \label{fig:Splitting}
\end{figure}

\begin{figure}[t]
  \centering
  \includegraphics[width=\linewidth]{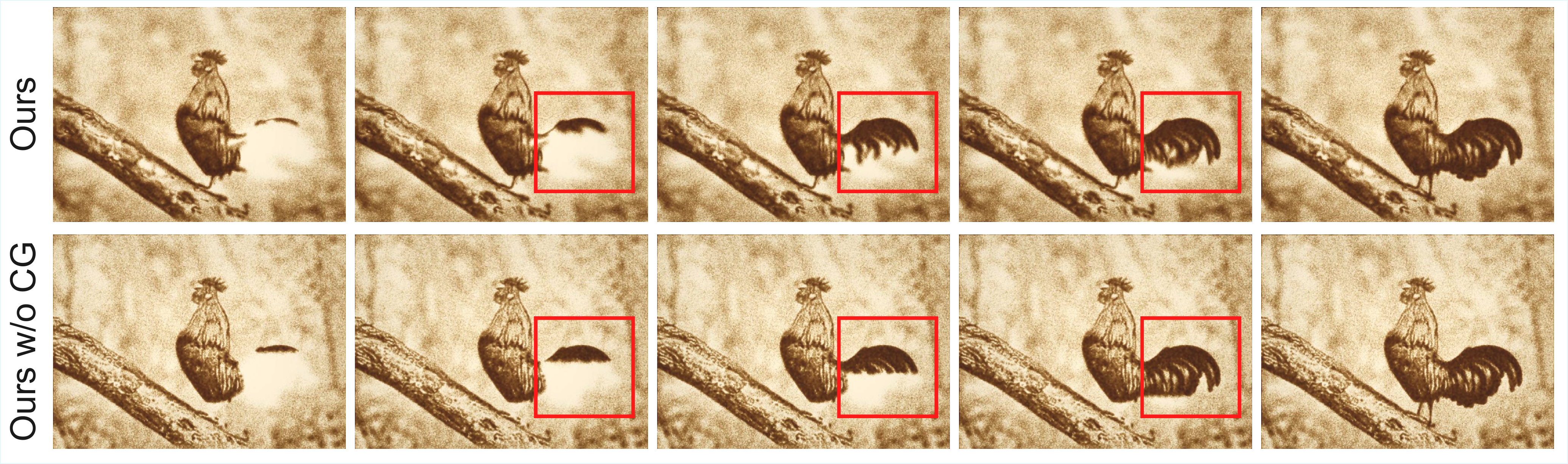}
  \caption{\textbf{Ablation of Curve Guidance.} 
Without curve guidance, primitives fail to form coherent strokes, as seen in the rooster's tail. 
Our full method produces structured patterns.}

  \label{fig:Guidance}
\end{figure}

\begin{table}[htbp]
  \centering
  \caption{Quantitative comparison of different painting generation methods. The best results are highlighted in \textbf{bold}.}
  \label{tab:quantitative_evaluation}
  \begin{tabular}{l|ccccc}
    \toprule
    Method & LPIPS $\downarrow$ & SSIM $\uparrow$ & FID $\downarrow$ & DDC $\downarrow$ & GTC $\downarrow$ \\
    \midrule
    Inverse Painting & 0.510 & 0.594 & \textbf{116.1} & 1.296 & 0.197 \\
    Paint Transformer & 0.708 & 0.461 & 359.3 & 1.620 & 1.469 \\
    PaintsUndo      & 0.652 & 0.506 & 187.5 & 1.350 & 0.208 \\
    \midrule
    Ours             & \textbf{0.395} & \textbf{0.644} & 223.8 & \textbf{0.876} & \textbf{0.083} \\
    \bottomrule
  \end{tabular}
\end{table}

\subsection{Ablation Study}

We conduct ablation studies to evaluate the key components of our method, 
including (1) semantic planning (SP), (2) curve guidance (CG), 
(3) subtractive color rendering (SR), and (4) dynamic topology optimization, including stroke merging (SM) and stroke splitting (SS).
We use LPIPS and SSIM to measure the temporal consistency 
of the generated drawing process, PSNR to evaluate the final reconstruction quality, 
and the number of strokes (\#Strokes) to reflect representation efficiency. 
Table~\ref{tab:ablation_study} shows the comparison results.
(1) Removing SP has limited impact on the final reconstruction quality, but significantly degrades temporal consistency, indicating its importance in guiding a coherent and semantically meaningful drawing process.
(2) Without CG, primitives become disorganized and fail to form coherent stroke structures, resulting in unrealistic drawing trajectories. As shown in Fig.~\ref{fig:Guidance}, our full method produces more structured and stroke-like patterns.
(3) With other rendering parameters fixed, removing SR leads to consistent degradation across all metrics, demonstrating that physically consistent subtractive color blending is essential for faithful reconstruction.
(4) Removing SM increases the number of strokes, indicating redundant and fragmented structures, which reduces representation efficiency.
(5) Disabling SS reduces the flexibility of stroke representation, making it difficult to capture local geometric variations. As shown in Fig.~\ref{fig:Splitting}, this results in erroneous cross-region fitting and blurred structural boundaries.

\begin{table}[htbp]
  \centering
  \caption{Ablation study on different components of our proposed method.}
  \label{tab:ablation_study}
  \begin{tabular}{l|cccc}
    \hline
    Method & LPIPS $\downarrow$ & SSIM $\uparrow$ & PSNR $\uparrow$ & \#Strokes\\
    \hline
    Ours w/o SP & 0.606 & 0.490 & 24.68 & +0.7\% \\
    Ours w/o CG & 0.487 & 0.594 & \textbf{25.09} & -- \\
    Ours w/o SR & 0.498 & 0.527 & 22.89 & 0.0\% \\
    Ours w/o SM & 0.425 & 0.615 & 24.27 & +16.5\% \\
    Ours w/o SS & 0.413 & 0.629 & 24.18 & -2.7\% \\
    \hline
    Ours        & \textbf{0.395} & \textbf{0.644} & 24.76 & 0.0\% \\
    \hline
  \end{tabular}
\end{table}

\begin{table}[t]
\centering
\caption{User study on human-likeness and preference.}
\begin{tabular}{lcc}
\toprule
Comparison & Human-like (\%) & Preference (\%) \\
\midrule
Ours vs. Inverse Painting & 74.6 & 68.9 \\
Ours vs. PaintsUndo & 85.3 & 70.5 \\
Ours vs. Paint Transformer & 86.2 & 78.8 \\
\bottomrule
\end{tabular}
\label{tab:user_study}
\end{table}

\subsection{User Study}
We conduct a user study to evaluate the human-likeness and preference of the generated sand painting processes. The study involves 40 participants, each presented with 18 pairs of painting sequences comparing our method with baseline approaches.
For each pair, participants are asked to select (1) which sequence better resembles a human sand painting process, and (2) which result they prefer overall. Each sequence is presented as a short video clip with a fixed duration and consistent playback settings.
The results, summarized in Table~\ref{tab:user_study}, show that our method is consistently preferred and is more frequently regarded as human-like compared to the baselines, indicating that our approach better captures realistic and plausible sand painting dynamics.

\section{Conclusion}

We presented SandSim, a framework for reconstructing sand painting processes from a single image. Our method uses curve-guided Gaussian splatting to model strokes as continuous Gaussian sequences, enabling structured sand deposition in a differentiable manner. Combined with semantic planning, the system decomposes scenes into ordered regions and generates temporally coherent drawing sequences. Experiments show that SandSim produces realistic sand painting processes with coherent stroke growth and plausible sand accumulation.

\textbf{Limitation and future work.}
Our framework mainly models sand deposition and does not explicitly simulate wiping or erasing operations, which limits its ability to reproduce removal effects and fine-grained corrections observed in real sand painting. In addition, our semantic planning relies on upstream models, making it susceptible to segmentation errors and ambiguities in structurally complex scenes, especially when semantic boundaries are unclear. Future work will focus on developing more robust planning mechanisms and more interactive, controllable sand painting systems, including text-guided generation via natural language.


\bibliographystyle{ACM-Reference-Format}
\bibliography{refs}

\end{document}